\documentclass[aps,pra,twocolumn,nofootinbib,preprintnumbers,english,showpacs]{revtex4-1}
\usepackage{amsthm,amsmath,amsfonts,dsfont,upgreek}    
\usepackage{amssymb,epsfig,setspace}
\usepackage{graphicx}
\usepackage{babel}
\usepackage{verbatim}

\begin{document}

\hyphenation{ano-ther ge-ne-ra-te dif-fe-rent know-le-d-ge po-ly-no-mi-al}
\hyphenation{me-di-um  or-tho-go-nal as-su-ming pri-mi-ti-ve pe-ri-o-di-ci-ty}
\hyphenation{mul-ti-p-le-sca-t-te-ri-ng i-te-ra-ti-ng e-q-ua-ti-on}
\hyphenation{wa-ves di-men-si-o-nal ge-ne-ral the-o-ry sca-t-te-ri-ng}
\hyphenation{di-f-fe-r-ent tra-je-c-to-ries e-le-c-tro-ma-g-ne-tic pho-to-nic}
\hyphenation{Ray-le-i-gh di-n-ger Kra-jew-ska Wal-czak Ham-bur-ger Ad-di-ti-o-nal-ly}
\hyphenation{Kon-ver-genz-the-o-rie ori-gi-nal in-vi-si-b-le cha-rac-te-ri-zed}
\hyphenation{Ne-ver-the-less sa-tu-ra-te Ene-r-gy sa-ti-s-fy le-vels re-s-pec-ti-ve pro-pe-r-ty}
\hyphenation{dif-fe-rent no-men-cla-tu-re re-gar-ding}

\title{Quantum models with spectrum generated by the flows of polynomial zeros}

\author{Alexander Moroz} 

\affiliation{Wave-scattering.com} 
 
\begin{abstract}
\noindent A class ${\cal R}$ of purely bosonic models is characterized 
having the following properties in a Hilbert space 
of analytic functions: 
(i) wave function 
$\uppsi(\upepsilon,z)=\sum_{n=0}^\infty \phi_n(\upepsilon) z^n$ 
is the {\em generating function} 
for orthogonal polynomials $\phi_n(\upepsilon)$ of 
a {\em discrete} energy variable $\upepsilon$, 
(ii) 
any Hamiltonian $\hat{H}_b\in {\cal R}$ 
has nondegenerate purely point spectrum that corresponds 
to infinite discrete support of measure
$d\nu(x)$ in the orthogonality relation of the polynomials 
$\phi_n$,
(iii) the support is determined exclusively by
the points of discontinuity of $\nu(x)$,
(iv) the spectrum of $\hat{H}_b\in {\cal R}$ can be numerically
determined as fixed points of monotonic flows of the zeros of 
orthogonal polynomials $\phi_n(\upepsilon)$,
(v) one can compute practically an unlimited number of 
energy levels (e.g. $2^{53}$ in double precision).
If a model of ${\cal R}$ is exactly solvable, its spectrum 
can only assume one of four qualitatively different types.
The results are applied to spin-boson quantum models 
that are, at least partially, diagonalizable and have at least single
one-dimensional irreducible component in the spin 
subspace. Examples include the Rabi model 
and its various generalizations. 
\end{abstract}


\maketitle

\section{Introduction}
\label{sc:intr}
Our work concerns models for which one can formulate 
a formal quantization criterion in terms of infinite continued fraction 
\begin{equation}
\mathbb{F}(x)\equiv a_0 + \frac{-b_{1}}{a_{1}-} \frac{b_{2}}{a_{2}-}
\frac{b_{3}}{a_{3}-}\cdots =0,
\label{fdfs}
\end{equation}
where the coefficients $a_n$ and $b_n\ne 0$ are functions 
of an energy variable $x$.
A prototype of the criterion (\ref{fdfs}) is the 
{\em Schweber quantization condition} (cf. Eq. (A.16) of Ref. \cite{Schw}) 
initially formulated for
a displaced harmonic oscillator and the Rabi, or single boson, model \cite{Rb}.
The gist of the present work is to explore consequences of that 
the {\em zeros} of the function $\mathbb{F}(x)$ 
are equivalent to the {\em poles} of infinite continued fraction
\begin{equation}
\mathbb{E}(x)\equiv \frac{-b_{0}}{a_{0}-} \frac{b_{1}}{a_{1}-}
\frac{b_{2}}{a_{2}-}\cdots = -\frac{b_{0}}{\mathbb{F}(x)}\cdot
\label{ttdf}
\end{equation}

In particular, let us consider the Schr\"{o}dinger  equation  
\begin{equation}
\hat{H}\Uppsi(z) =E \Uppsi(z)
\label{eme}
\end{equation}
induced by a model Hamiltonian $\hat{H}$ in the product Hilbert space 
${\cal B}=\mathfrak{b}\otimes\mathbb{C}^N$, where 
$\mathfrak{b}$ is the Bargmann space of analytic functions 
 \cite{Schw,Brg} and $\mathbb{C}^N$ is 
$N$-dimensional spin subspace \cite{Br,Al,AMep,AMops,AMtb,Zh3}. 
We assume that $\hat{H}$
possesses a symmetry group $G$ that has at least single
{\em one-dimensional irreducible representation} in the spin 
subspace \cite{Br,Al,AMep,AMops,AMtb,Zh3,ZhN,SS}. 
On many occasions (e.g. if $\hat{H}$ is a linear combination of 
the bosonic operators $a^+a$, $a$ and $a^+$) 
the purely bosonic Hamiltonian $\hat{H}_b$ describing  
the irreducible component is intrinsically {\em tridiagonal}.
Then the eigenvalue equation reduces to 
a {\em three-term recurrence relation} (TTRR) \cite{Schw,AMep,AMops,AMtb,Zh3}
\begin{equation}
\phi_{n+1} + a_n \phi_n + b_n \phi_{n-1}=0 \hspace*{1.8cm}  (n\ge 1),
\label{3trg}
\end{equation}
and a {\em two-term} condition on $\phi_0$ and $\phi_1$ \cite{AMep},
\begin{equation}
\phi_{1} + a_0 \phi_0 =0.
\label{2tr}
\end{equation}
Here $\{\phi_n\}_{n=0}^\infty$ are the sought 
expansion coefficients of an entire function in $\mathfrak{b}$,
\begin{equation}
\uppsi(z)=\sum_{n=0}^\infty \phi_n z^n,
\label{pss}
\end{equation}
corresponding to the one-dimensional irreducible component 
of $\Uppsi$ described by $\hat{H}_b$ (see Sec. \ref{sec:aes} below).
In fact, there always exists an orthonormal basis $\{{\bf e}_n\}_{n=0}^\infty$ 
such that a given self-adjoint operator $\hat{H}$ takes on a {\em tridiagonal} form,
\begin{equation}
\hat{H}{\bf e}_n = \tilde{a}_n{\bf e}_n + \tilde{b}_{n+1}{\bf e}_{n+1} + \tilde{b}_n{\bf e}_{n-1},
\label{hd4p1}
\end{equation}
with {\em real} recurrence coefficients and with $\tilde{b}_n\ge 0$, $n\ge 0$ \cite{Hd,AMhd}.
In such a basis,
(i) the expansion coefficients $\phi_n$ are polynomials of the 
$n$-th order of an {\em orthogonal polynomial sequence} (OPS) 
of a {\em discrete} variable \cite{AMops,AMtb,Hd,AMhd,NSU,Lor}
and hence
(ii) the wave function $\uppsi(z)$ is the {\em generating function} 
for the polynomials \cite{AMops,AMtb,Hd,AMhd}.
Here the discrete variable means that 
the {\em distribution} function $\nu(x)$ in the orthogonality 
relations of the polynomials
(see appendix) is an increasing {\em step} function.

The outline of the present work is as follows.
In Sec. \ref{sec:nta} we define a {\em recurrence class of 
purely bosonic models}, ${\cal R}$,
for which the quantization criterion (\ref{fdfs}) can be shown to
follow from (\ref{3trg}) \cite{Schw,AMep,AMops,AMtb,Zh3}. 
The class ${\cal R}$ is broad enough to encompass 
the Rabi model \cite{AMops,AMtb} and its various generalizations. 
The Rabi model, which describes the simplest fully quantized interaction 
between light and matter [cf. Eq. (\ref{rabih}) below], can be realized 
in a rich variety of different setups such as Josephson junctions,
circuit quantum electrodynamics, trapped ions, superconductors,
and semiconductors \cite{KGK,BGA,FLM,NDH}.
The model plays a fundamental role in various applications 
of quantum optics, in implementation of 
diverse protocols in contemporary quantum 
information, with potential applications to future 
quantum technologies \cite{KGK,BGA,FLM,NDH}. 
In its semiclassical form, the model is the basis 
for understanding nuclear magnetic resonance.

In Sec. \ref{sec:proof} it is proven that 
$\mathbb{E}(x)$ is related to the measure $d\nu(x)$ 
by the {\em Stieltjes transform} [Eq. (\ref{erpt}) below].
$\mathbb{E}(x)$ has only simple poles for $\hat{H}_b\in {\cal R}$,
which coincide with the points of discontinuity of $\nu(x)$. 
On labeling the poles of $\mathbb{E}(x)$ in increasing order, 
the $n$th pole can be alternatively recovered as a fixed 
point of the flow of the $n$th zeros of polynomials of the OPS.
The transformation (\ref{ttdf}) from $\mathbb{F}(x)$ to $\mathbb{E}(x)$ 
enables one to translate the above results for 
the {\em poles} of $\mathbb{E}(x)$ to those for the 
{\em zeros} of $\mathbb{F}(x)$. 
In particular, $\hat{H}_b\in {\cal R}$
has nondegenerate purely point spectrum \cite{cc} that
corresponds to the points of discontinuity of $\nu(x)$.
As elaborated in Sec. \ref{sec:num}, a direct 
practical consequence of the result is an entirely 
new, efficient, and relatively general method 
in determining the spectrum. 
Contrary to searching for zeros of $\mathbb{F}(x)$, 
the spectrum coincides with the limit points of
the flows of zeros of appropriate orthogonal polynomials 
of a discrete variable,
which can be determined much more efficiently.
For example, on using a very simple stepping algorithm,  
we were able to determine up to ca {\em 1350} 
energy levels per parity subspace for the Rabi model 
[i.e. almost two orders of 
magnitude more than is possible to obtain from the 
Schweber quantization condition (\ref{fdfs})] \cite{AMtb}. 
We sketch the basic features of an improved algorithm that allows 
to determine practically an {\em unlimited} number of energy levels 
within corresponding machine precision.

Sec. \ref{sec:disc} is divided into a number of subsections
where our results are extensively discussed from various angles:
comparison of ${\cal R}$ and quasi-exactly-solvable models 
(sec. \ref{sec:dgn}), 
the concept of {\em almost exactly solvable models} 
(sec. \ref{sec:aes}), numerical issues (sec. \ref{sec:dnm}),
and relation with earlier work (sec. \ref{sec:drl}).
We then conclude with Sec. \ref{sec:conc}.
Some additional technical remarks are relegated to
appendices.

\section{Recurrence class ${\cal R}$ of purely bosonic 
models}
\label{sec:nta}
The coefficients $\{\phi_n\}_{n=0}^\infty$ define 
an entire function in $\mathfrak{b}$ whenever the sum
$\sum_{n=0}^\infty |\phi_n|^2 n!$
converges (cf. Eq. (1.4) of Ref. \cite{Brg}). (Note in passing
that the Bargmann condition, which corresponds to the 
Hilbert space of entire functions of growth 
$(\frac{1}{2}, 2)$, presumes the standard measure
$(1/\pi)e^{-|z|^2}\, dzd\bar{z}$ in $\mathbb{C}$. 
In the Hilbert spaces of entire functions of different growth
another measure and convergence criterion apply \cite{BG}.)
The quantization criterion (\ref{fdfs}) 
is rigorous consequence of the eigenvalue equation 
(\ref{eme}) provided that the TTRR (\ref{3trg}) 
[unless otherwise stated, considered in the absence of the 
two-term condition (\ref{2tr})] has a {\em minimal} 
solution $\{m_n\}_{n=0}^\infty$ with $m_0\ne 0$ 
(cf. Theorem 1.1 due to Pincherle in Ref. \cite{Gt}).
The minimal solution exists if for any other linearly 
independent solution $\{d_n\}_{n=0}^\infty$ 
of the TTRR (\ref{3trg}) one has $\lim_{n\rightarrow\infty} m_n/d_n =0$ \cite{Gt}. 
In the latter case all other linearly independent solutions 
are called {\em dominant} \cite{Gt}. The latter are not unique as 
any linear combination of $m_n$ and $d_n$ yields another dominant solution.
(Note that there might be TTRR which do not have any 
minimal solution \cite{Gtp}.)

In what follows, we limit ourselves further to the case when:\newline

({\bf A}) the coefficients $a_n$'s are {\em linear} functions 
of an energy variable $x$, i.e., $a_n=-(\alpha_n x - c_n)$, where 
the coefficients $\alpha_n$ and $c_n$  
are {\em real} and independent of $x$, 
and $\alpha_n\ne 0$ for $n\ge 1$.\newline

\noindent The condition ({\bf A}) is not a serious constraint, because it is 
always met in Haydock's basis \cite{Hd}.
The TTRR (\ref{3trg}) then becomes a defining equation for 
{\em orthogonal polynomials}.
Indeed, according to the {\em Favard-Shohat-Natanson} theorems 
(given as 
Theorems I-4.1 and I-4.4 of Ref. \cite{Chi}),
the necessary and sufficient condition for a family of 
polynomials $\{p_n\}$ (with degree $p_n = n$) to 
form a positive definite OPS is that $p_n$'s satisfy 
a TTRR (\ref{3trg}) and (\ref{2tr}) 
with the coefficients as specified above, 
together with the initial condition $p_{-1} = 0$ and $p_{0}=$ const.
Without any loss of generality, a suitable rescaling of 
$\phi_n$ \cite{AMops,AMtb} enables one to recast (\ref{3trg}) 
as the TTRR of a {\em monic} OPS \cite{ntt}
\begin{eqnarray}
p_n(x) &=&(x-c_{n-1})p_{n-1}(x)- \lambda_{n-1} p_{n-2}(x),
\label{chi3tr}
\\
p_{-1}(x) &=&  0,~~~~~~ p_{0}(x)=1,
\label{chi3tric}
\end{eqnarray}
where $\lambda_n\ne 0$, $n\ge 1$ and we keep the notation
$c_n$ also for the rescaled coefficients. 
Because neither the TTRR (\ref{3trg}) nor the two-term 
condition (\ref{2tr}) contains $b_0$, one can in virtue of 
$p_{-1}\equiv 0$ always set the rescaled $b_0$ 
as $\lambda_{0}=1$ \cite{AMops,AMtb}.

Because the present work is concerned with the part of spectrum 
corresponding to a {\em one-dimensional irreducible component} 
of $\Uppsi$ described by purely bosonic $\hat{H}_b\in {\cal R}$, 
it is obvious to focus 
on the {\em discrete} spectrum. The spectrum can be discrete only if 
(i) dominant solutions 
of the TTRR (\ref{3trg}) do not generate an element of $\mathfrak{b}$ 
and, simultaneously, 
(ii) the minimal solution of the TTRR (\ref{3trg}) does so.
Indeed, the system of the TTRR (\ref{3trg}) with (\ref{2tr}) 
as an initial condition has always a {\em unique} solution \cite{Gt}. 
The unique solution is in general 
a linear combination of the minimal and dominant solutions.
However, the unique solution is in $\mathfrak{b}$ only 
in the special case if it reduces to the minimal solution \cite{AMep}.
Under all other circumstances either both minimal and 
dominant solutions generate
functions from $\mathfrak{b}$, or none does so.
In the latter case the spectrum is obviously {\em empty}, 
whereas in the former case the spectrum is 
necessarily {\em continuous}. Indeed, for any energy there would exist
a unique solution of the TTRR (\ref{3trg}) with (\ref{2tr}) \cite{Gt}.
Irrespective if the solution is a dominant or minimal one, 
it would be in $\mathfrak{b}$, and hence in the spectrum.

Let the recurrence coefficients assume 
an asymptotic powerlike dependence as a function of $n$ \cite{AMep}
\begin{equation}
a_n\sim a n^{\alpha},~~~~~~~~~~~ 
          b_n\sim b n^{\beta}\hspace*{1.2cm} (n\rightarrow\infty),
\label{rcd}
\end{equation}
where $a$ and $b$ are proportionality constants.
The {\em Perron-Kreuser}  theorem (Theorem 2.3 in Ref. \cite{Gt}) implies that
(i) a {\em minimal} solution exists and (ii) {\em discrete} spectrum 
is possible provided that any of the following alternatives is satisfied:
\begin{itemize}

\item {\bf (a)} $\alpha>-1/2$, $\beta<\alpha-(1/2)$,

\item {\bf (b)} $\alpha>-1/2$, $\beta=\alpha-(1/2)$, $|b|<|a|$,

\item {\bf (c)} $\alpha=-1/2$, $|a|\ge 1$, $\beta<-1$,

\item {\bf (d)} $\alpha=-1/2$, $\beta=-1$, $|t_1|\ge 1$, $|t_2|<1$.

\end{itemize}
Here the first three alternatives follow from case (a) of 
the Perron-Kreuser theorem,
whereas the last one is a consequence of the theorem case (b).
The conditions {\bf (a)}-{\bf (d)} define a ``{\em recurrence}" class ${\cal R}$
of quantum models. The class is here defined broader than in our earlier 
work \cite{AMep} by including also cases {\bf (c)} and {\bf (d)},
in order to accommodate more general 
Hilbert spaces of entire functions \cite{BG}.

\subsection{Examples} 
\label{sec:ex}
After a rather formal and abstract introduction of ${\cal R}$, we argue 
that the conditions for ${\cal R}$ are, in broad sense, natural. For example, 
the condition ({\bf A}) is automatically satisfied 
if $\hat{H}$ is a {\em linear} combination 
of $a^+a$, $a$ and $a^+$. 
Indeed, upon the action of $a^+a$, $a=(d/dz)$ 
and $a^+=z$ on $\uppsi(z)$ in Eq. (\ref{pss}),
the coefficient of the resulting $z^n$ monomial will become $n\phi_n$, 
$(n+1)\phi_{n+1}$, and  $\phi_{n-1}$, respectively.
Therefore, if $\hat{H}$ is a linear combination of $a^+a$, $a$ and $a^+$, 
the eigenvalue equation for any $\hat{H}_b\in {\cal R}$  describing
the one-dimensional 
irreducible component of $\Uppsi$ inevitably reduces to 
a {\em three-term recurrence relation} (TTRR) of the type ({\bf A}) 
with $\alpha=0$ and $\beta=-1$, i.e. corresponding to case {\bf (a)}.
Even if the condition ({\bf A}) is not automatically satisfied,
it is only a question of finding an appropriate orthonormal basis $\{{\bf e}_n\}_{n=0}^\infty$ 
to bring a given Hamiltonian to a tridiagonal form (\ref{hd4p1}) \cite{Hd,AMhd}.

Not surprisingly, the class ${\cal R}$ is broad enough to encompass
the Rabi model \cite{AMops,AMtb} and its various generalizations. 
The Rabi model \cite{Rb} describes the 
simplest interaction between a 
cavity mode with a frequency $\omega$ and a two-level system 
with a resonance frequency $\omega_0$. 
The model is characterized by the Hamiltonian \cite{Schw,Rb} 
\begin{equation}
\hat{H}_R =
\hbar \omega \mathds{1} \hat{a}^\dagger \hat{a}  
 +  \hbar g\sigma_1 (\hat{a}^\dagger + \hat{a}) + \mu \sigma_3
\label{rabih}
\end{equation}
acting in the Hilbert space ${\cal B}=\mathfrak{b}\otimes\mathbb{C}^2$,
where $\mu=\hbar \omega_0/2$, $\hat{a}$ and $\hat{a}^\dagger$ are 
the conventional boson annihilation and creation operators 
satisfying commutation relation $[\hat{a},\hat{a}^{\dagger}] = 1$,
and $g$ is a coupling constant \cite{Schw,Brg}. 
In what follows, $\mathds{1}$ is the unit matrix,
$\sigma_j$ are the Pauli matrices in 
their standard representation, and we set the 
reduced Planck constant $\hbar=1$.
$\hat{H}_R$ is invariant under the parity 
$\hat{\Pi}=\sigma_3 \hat{\pi}$, where
$\hat{\pi}a\hat{\pi}^{-1}=-a$ 
and $\hat{\pi}a^+\hat{\pi}^{-1}=-a^+$.
${\cal B}$ can be thus written as a direct sum 
${\cal B}={\cal B}_+\oplus {\cal B}_-$ 
of the parity eigenspaces, or of invariant subspaces ${\cal B}_\pm$.
In each of them the Rabi model is characterized 
by a corresponding three-term recurrence (cf. Eq. (37) of Ref. \cite{AMep})
\begin{eqnarray}
 \lefteqn{
 \phi_{n+1}^\pm +\frac{1}{\kappa (n+1)}\, 
              [n  - \upepsilon  \pm(-1)^n\Delta]\phi_{n}^\pm 
}\hspace*{3cm}
\nonumber\\
      &&    + \frac{1}{n+1}\, \phi_{n-1}^\pm = 0,
\label{rbmb2}
\end{eqnarray}
where energy variable $x=\upepsilon\equiv E^\pm/\omega$, 
$\kappa=g/\omega$ reflects the coupling strength, and
$\Delta=\mu/\omega$ \cite{AMep}. Upon comparing with 
(\ref{rcd}), one has $\alpha=0$, $\beta=-1<\alpha-(1/2)$, 
which corresponds to case {\bf (a)} \cite{AMep,AMops,AMtb}. 
The substitution $\phi_n\rightarrow P_n/n!$
transforms the initial recurrence (\ref{rbmb2}) 
into TTRR (\ref{chi3tr}) of a positive definite monic OPS \cite{AMops}.
The case of a displaced harmonic oscillator is the 
exactly solvable limit of $\hat{H}_R$ for $\mu=0$ that corresponds to
$\Delta=0$, whereby the recurrences (\ref{rbmb2}) reduce 
to Eq. (A.17) of Ref. \cite{Schw}.

It is useful to remind here that the Rabi model with
a ``wrong" negative sign of its parameters $g$ and $\mu$ 
(cf. Eq. 12 of Ref. \cite{SS})  
was used to describe an excitation hopping between two sites and 
the interaction of a dipolar impurity (paraelectric or paraelastic) 
with a crystal lattice \cite{SS}. The sign change 
induces sign reversal of $\kappa$ and $\Delta$ in the TTRR (\ref{rbmb2}),
but otherwise does not change any its essential features.

Further models can be obtained by changing or introducing different interaction
terms to the Rabi model. Following Appendix of Ref. \cite{Wg},
the Rabi Hamiltonian (\ref{rabih}) supplemented with a
{\em momentum} dependent interaction term
\begin{equation}
V=i\sigma_2 g_b(a - a^+)
\nonumber 
\end{equation}
remains invariant with regard to the parity operator
$\hat{\Pi}$, and thus amenable to the 
Fulton-Gouterman transformation (FGT) \cite{Br,FG}, 
resulting in a pair of TTRR. 

Several groups  \cite{SBB,TAP} studied the generalized Rabi model
\begin{equation}
\hat{H}_{\mathrm{gR}}  = \omega \hat{a}^\dagger \hat{a}  +  \mu \sigma_3
 + g_1 ( \hat{a}^\dagger \sigma_- + a\sigma_+)  +
   g_2 (\hat{a}^\dagger \sigma_+  + a\sigma_-),
\nonumber 
\end{equation}
which interpolates between the Jaynes and Cummings 
model (for $g_2=0$) and the original Rabi model $g_1=g_2$. 
The model, which is again invariant with regard to the parity operator
$\hat{\Pi}$, and thus amenable to the FGT \cite{FG}, 
can be mapped onto the model describing a two-dimensional
electron gas with {\em Rashba} ($g_R\sim g_1$) 
and {\em Dresselhaus} ($g_D\sim g_2$) 
spin-orbit couplings subject to a perpendicular magnetic field.

There is an outside chance that a {\em driven Rabi model} having an extra
driving term $\lambda\sigma_1$ \cite{Zh3,GD,La}
could also be treated within our framework.
Although the driven Rabi model is not invariant with regard to $\hat{\Pi}$,
Gardas and Dajka \cite{GD} argued that it
possesses a {\em nonlocal} parity. However explicit 
form of the {\em nonlocal} parity
operator has not been provided and it is unclear 
if it could be useful for a FGT.

Another option is to consider models in the spin space $\mathbb{C}^N$
with $N>2$. The so-called Rabi model for $N$-state atoms 
which can be diagonalized in the spin subspace has been recently studied 
by Albert \cite{Al}. A $\mathbb{Z}_N$ symmetric chiral Rabi model has been 
recently introduced by Zhang \cite{ZhN}.

{\em Nonlinear} single-mode terms (such as $a^k$ and $(a^+)^k$ with $k\ge 2$)
and multi-mode terms (e.g. $a_1a_2$ and $a_1^+a_2^+$) would naively lead
to higher-order recurrences than the fundamental TTRR (\ref{3trg}). 
Nonetheless, through judicious application of the representation theory 
for higher order  polynomial deformations of the $su(1,1)$ Lie algebra via 
a Jordan-Schwinger like construction method \cite{Zh3,LYZ1,LYZ}, and 
ensuing algebraization of the spin-boson systems, such {\em nonlinear}
models are not excluded from the scope of the present work.
Indeed, as shown by Zhang \cite{Zh3}, a TTRR also arises in the case 
of the {\em two-photon} and {\em two-mode Rabi models}, which both enjoy a
parity symmetry and which can be described by $\hat{H}_b\in {\cal R}$ \cite{ZhM}.

\section{Spectrum generated by the flows of polynomial zeros}
\label{sec:proof}
For $\hat{H}_b\in {\cal R}$, the infinite continued fraction
in Eq. (\ref{fdfs}) can be expressed as the limit \cite{AMops}
\begin{equation}
r_0=\lim_{n\rightarrow\infty} \frac{P_{n-1}^{(2)}(x)}{P_n^{(1)}(x)},
\label{r0lo}
\end{equation}
where, given TTRR (\ref{chi3tr}), the polynomials
$P_{n}^{(\upsilon)}$, $\upsilon=0,1,2$ are defined by \cite{asc}
\begin{equation}
P_n^{(\upsilon)}(x)= (x-c_{n-1+\upsilon})P_{n-1}^{(\upsilon)}(x)
            - \lambda_{n-1+\upsilon} P_{n-2}^{(\upsilon)}(x),
              ~~~(n\ge 1).
\label{chi3tra}
\end{equation}
The initial condition is the same as in Eq. (\ref{chi3tric}).
For the sake of notation,
the polynomials of the OPS for $\upsilon=0$ will be 
denoted simply as $P_n$.
Thus the $p_n$'s, defined earlier by the TTRR (\ref{chi3tr}) 
that follows directly from the initial TTRR (\ref{3trg}), 
has become $P_n$.
The respective monic OPS with $\upsilon=1,2$ are called 
{\em associated} to $\{P_n\}$ (see Sec. III-4 of Ref. \cite{Chi}) \cite{asc}.
The need of three different OPS
is obvious: whereas $\{P_n\}$ determines the expansion
coefficients of the physical state, the pair 
$\{P_n^{(1)}\}$ and $\{P_n^{(2)}\}$ defines the infinite continued fraction
in Eq. (\ref{fdfs}).

Analogously to the infinite continued fraction in Eq. (\ref{fdfs}), 
\begin{equation}
\mathbb{E}(x)=\lim_{n\rightarrow\infty} \frac{P_{n-1}^{(1)}(x)}{P_n(x)}
\label{Exe}
\end{equation}
[cf. Eq. (\ref{r0lo})].
Indeed, the infinite continued fraction (\ref{ttdf}) is obtained from that
in Eq. (\ref{fdfs}) by a substitution $(a_n,b_n)\rightarrow (a_{n-1},b_{n-1})$.
The latter corresponds to the substitution 
$\upsilon\rightarrow \upsilon-1$ in Eq. (\ref{chi3tra}).

Let $x_{nl}$, $l=1,2,\ldots,n$, denote the zeros 
of $P_n(x)$ arranged in increasing order \cite{pnz}. 
For any $n$ and $l=1,2,\ldots, n-1$ one has
\begin{equation}
x_{nl}<x_{n-1,l}<x_{n,l+1}
\label{1p5p4}
\end{equation}
(cf. Theorem I-5.3 of Ref. \cite{Chi}).
Because $x_{nl}<x_{n,l+1}$, the zeros of any $P_n(x)$ 
are all {\em simple} (cf. Theorem I-5.2 of Ref. \cite{Chi}). 
The first inequality in (\ref{1p5p4}) implies that the sequence
$\{x_{nl}\}_{n=l}^\infty$ is strictly {\em decreasing}
for any fixed $l$. Therefore, the respective limits 
\begin{equation}
\lim_{n\rightarrow\infty} x_{nl} = \xi_l
\label{1p5p6}
\end{equation}
exist. The above properties are intrinsic signatures of any OPS \cite{Chi}.
In what follows, we denote the set of all the limit points by
$\Xi=\{\xi_l\,| \, l=1,2,3,\ldots\}$.

Corresponding to the OPS $\{P_n\}$, there is
a positive-definite moment functional ${\cal L}$ (see Appendix \ref{app:ttrr}).
According to the {\em representation theorem} 
(Theorem II-3.1 of Ref. \cite{Chi}), ${\cal L}$ can be characterized by 
a right continuous distribution function $\nu$
that is determined through a suitable limit process (see Appendix \ref{app:ttrr}).
The set of all the points $x$ where $\nu(x)$ has either a finite
jump or increases continuously,
\begin{equation}
\mathfrak{S}(\nu)=\{x\,| \, 
\nu(x+\delta) - \nu(x-\delta)>0 \mbox{ for all }\delta>0 \},
\nonumber
\end{equation}
is called the spectrum of $\nu$, or alternatively 
the support of ${\cal L}$ (cf. p. 51 of Ref. \cite{Chi}), 
or the support of the Stieltjes integral measure $d\nu$ induced by $\nu$.
For any positive-definite moment functional the set is {\em infinite} \cite{Chi}.

On recalling the arguments of Ref. \cite{AMops},
$\mathbb{E}(z)$ can be defined as a {\em regular analytic} 
function of a complex variable $z\in\mathbb{C}$,
\begin{equation}
\mathbb{E}(z) = \int_{-\infty}^\infty \frac{d\nu(x)}{z-x}\cdot
\label{erpt}
\end{equation}
$\mathbb{E}(z)$ is thus the {\em Stieltjes function} \cite{Ma88,Mg1,Wt}.
The distribution function $\nu$ in the 
{\em Stieltjes transform} representation (\ref{erpt}) is, 
assuming the normalization $\nu(-\infty)=0$, {\em unique}
(see footnote 30 on p. 268 of Ref. \cite{Hb1}).
The determinacy of the Stieltjes measure $d\nu$ 
follows also independently from Carleman's 
criterion (cf. Eq. (VI-1.14) of Ref. \cite{Chi}; p. 59 of Ref. \cite{ST}) 
which says that the moment problem is determined if  
$\sum_{l=1}^\infty \lambda_l^{-1/2} =\infty$.
The latter is obviously satisfied in our case.
Note that the polynomials on the r.h.s of Eq. (\ref{Exe}) 
are monic and all their zeros are on the real axis. 
Then if $\lim_{z\rightarrow\infty} z\mathbb{E}(z)$ exists 
and if the limit is finite (e.g. equals to one) 
in any sector $\epsilon\le \mbox{arg } z\le \pi-\epsilon$, $0<\epsilon< \pi/2$,
the {\em Stieltjes transform} representation (\ref{erpt})
holds with a {\em bounded} and non-decreasing $\nu(x)$ 
(cf. Lemma 2.2 of Ref. \cite{ST}).

Hamburger's Theorem XII' \cite{Hb1} guarantees the 
{\em Stieltjes transform} representation (\ref{erpt})
in any closed finite region $\Omega$ 
of the complex plane $\mathbb{C}$ which does not contain any part of the real axis.
An important result of Ref. \cite{AMops} was that the representation (\ref{erpt})
can be extended to any closed interval on real axis located
within the open intervals where $\nu=$ const.
In other words, one can employ the representation (\ref{erpt}) 
within any closed interval
of the real axis which does not have any common point with $\mathfrak{S}(\nu)$.
The result can be regarded as an extension of the 
{\em Markov} theorem (Theorem 2.6.2 of Ismail book \cite{Isb} 
or p. 90 of Ref. \cite{Chi}).

Denote 
\begin{equation}
\sigma \equiv \lim_{j\rightarrow \infty} \xi_j.
\nonumber 
\end{equation}
There are only the following possibilities regarding 
the behavior of the $\xi_l$ (pp. 62-63 of Ref. \cite{Chi}):
\begin{itemize}

\item {\bf (a')} $\xi_l=\sigma=-\infty~~~~~~$  $(l\ge 1)$

\item {\bf (b')} $-\infty <\xi_1<\xi_2< \ldots <\xi_l =\sigma$ for some $l\ge 1$

\item {\bf (c')} $-\infty <\xi_1<\xi_2< \ldots <\xi_l
        < \ldots <\sigma=\infty$.

\end{itemize}
In the first two cases the set $\Xi$ is {\em finite} and thus cannot
coincide with the {\em infinite} $\mathfrak{S}(\nu)$. The latter comprises 
infinitely many points in addition to the elements of $\Xi$.
The first two cases are also unphysical. 
The first one already from the very fact that physical
models have their energy spectrum bounded from below, whereas
all energy levels would be at $-\infty$ in case {\bf (a')}.
In case {\bf (b')} the corresponding physical model would have,
following the analysis of Ref. \cite{AMops}, a finite number
of energy levels, with {\em infinitely degenerate} highest energy level.
We recall that energy levels corresponds to the zeros of $\mathbb{F}(z)$, which
are bracketed by the poles of $\mathbb{F}(z)$ \cite{AMops}.
According to Eq. (\ref{r0lo}), the poles of $\mathbb{F}(z)$ correspond to 
the limit points $\xi_l^{(1)}$ of sequences $\{x_{nl}^{(1)}\}_{n=l}^\infty$ \cite{AMops}.
Now for any associated OPS's 
(see sec. III.4 of Ref. \cite{Chi}), the zeros of $P_{n}^{(\upsilon)}(x)$ 
and $P_{n-1}^{(\upsilon+1)}(x)$ are {\em interlaced} 
(Theorem III-4.1 of Ref. \cite{Chi}). Specifically, 
\begin{equation}
x_{nl}^{(\upsilon)}<x_{n-1,l}^{(\upsilon+1)}<x_{n,l+1}^{(\upsilon)},
\hspace*{0.8cm} \upsilon=0,1.
\label{3p4p11}
\end{equation}
The latter implies $\xi_l\le \xi_l^{(1)} \le \xi_{l+1}$ and thereby
justifies the above conclusions for energy 
levels in cases {\bf (a')} and {\bf (b')}.

In case {\bf (c')}, the infinite spectrum $\mathfrak{S}(\nu)$ 
is formed exclusively by the points of $\Xi$ 
(see the summary of Sec. II-4 on pp. 62-63 of Ref. \cite{Chi}).
In other words, $\mathfrak{S}(\nu)$ reduces to a one-dimensional discrete lattice 
$\Lambda\equiv \Xi$ representing the infinite discrete support of $d\nu(x)$ \cite{cc}.
Indeed, $\nu$ experiences a {\em finite jump} at any point $\xi_k\in\Xi$,
\begin{equation}
0< \nu(\xi_k) - \nu(\xi_k-0) = {\cal M}_k = 
\left[\sum_{l=0}^\infty \frac{P_l^2(\xi_k)}{\mathfrak{n}_l}\right]^{-1},
\label{rsfrme}
\end{equation}
where 
\begin{equation}
\mathfrak{n}_l={\cal L}[1] \lambda_1\ldots \lambda_{l}=||P_l(x)||^2 
\nonumber 
\end{equation} 
is the squared norm of $P_l(x)$, and the positive numbers ${\cal M}_k$ 
satisfy the condition $\sum_{k=0}^\infty {\cal M}_k=1$ \cite{AMops,Chi}.
The determinacy of the Stieltjes measure $d\nu$
implies that at all other points of the real axis the sum in the square 
bracket is {\em divergent} (cf. Theorem 2.9 
and Corollary 2.8 of Ref. \cite{ST}; Theorem 2.5.3 
and Corollary 2.5.3 of Ref. \cite{Ak}). 
The divergence is a hallmark of that the TTRR (\ref{3trg}), with the
two-term condition (\ref{2tr}) taken as an initial condition, 
can only be satisfied by a {\em dominant} solution of the TTRR
\cite{AMep,Gt}.
Case {\bf (c')} implies that for any {\em physical} model of ${\cal R}$,  
the OPS defined by the TTRR (\ref{chi3tr}) have to be formed 
by the polynomials of a {\em discrete} variable with 
an unbounded spectrum $\mathfrak{S}(\nu)$.
Furthermore, in any irreducible subspace the model spectrum
is, as expected, {\em nondegenerate} \cite{AMops} 
(see Sec. \ref{sec:dgn} for discussion of this point).
There are no level crossings allowing the unique
labeling of each state.

On physical grounds we assume ${\cal R}$ 
be limited to case {\bf (c')} in what follows. 
Although there is a number of sufficient conditions 
on recurrence coefficients that ensure $\sigma=\infty$ 
(cf. Eq. (IV-3.7) of Ref. \cite{Chi}
that was employed for the Rabi model in Ref. \cite{AMops}),
they are expected to be satisfied for physical models and there is
no need to discuss them here.
Similarly to $\mathbb{F}(z)$ studied in Ref. \cite{AMops}, 
$\mathbb{E}(z)$ can be then represented as a {\em Mittag-Leffler} 
partial fraction decomposition,
\begin{equation}
\mathbb{E}(z) = \sum_{k=1}^\infty \frac{{\cal M}_k}{z-\xi_k},
\nonumber 
\end{equation}
defining a meromorphic function
in the complex plane $\mathbb{C}$ with {\em real simple} poles 
and {\em positive} residues.
The series is absolutely and uniformly 
convergent in any finite domain having a finite distance
from the simple poles $\xi_j$, 
and it defines there a holomorphic function of $z$.
The corresponding $\hat{H}_b\in {\cal R}$ 
(in general any one-dimensional 
irreducible component of $\Uppsi$ in a spin subspace - see Sec. \ref{sec:aes} below) 
has infinite number
of {\em nondegenerate} energy levels without (apart from $+\infty$) 
any accumulation point. 
This concludes the proof of the main result of the present work.

We have just shown that the distribution function $\nu(x)$ in the 
orthogonality relations of the polynomials 
of discrete variable is an increasing step function.
The spectrum of $\hat{H}_b\in {\cal R}$ 
corresponds to $\mathfrak{S}(\nu)$, which is 
given by the set $\Xi$ of points of discontinuity of $\nu(x)$.
Borrowing {\em renormalization group} (RG) language, 
the discrete flows generated by the polynomials zeros flow 
toward the spectral points. 
If $\Upsigma$ denotes the spectrum of $\hat{H}_b\in {\cal R}$,
$\Upsigma$ coincides with the corresponding 
discrete lattice $\Lambda\equiv \Xi = \mathfrak{S}(\nu)$.

\section{Numerical implications}
\label{sec:num}
Obviously, if one knows $\nu(x)$ in Eq. (\ref{erpt}) explicitly, 
one also knows $\mathfrak{S}(\nu)$, and
the corresponding model can be solved {\em exactly}. 
Unfortunately, a general procedure of recovering $d\nu(x)$ from 
an initial TTRR is not known. The task can only
be performed for the so-called {\em classical} OPS \cite{NSU,Lor,Chi,Isb}.
In all other cases, apart from some special cases \cite{Isb},
the spectrum have to be determined numerically.
However, one can identify numerically only a small number 
of the very first eigenvalues from the 
functional dependence of $\mathbb{F}(x)$ (cf. Figs. 1,2 of Ref. \cite{AMep}; 
Fig. 1 of Refs. \cite{AMops,AMcm}).
Soon afterwards, $\mathbb{F}(x)$ displays a {\em featureless} 
monotonically decreasing behavior \cite{AMtb} - cf. F77 code made 
available online \cite{AMr}.
The latter has been traced down to a curious property 
of zeros of associated OPS \cite{AMtb} - cf. data files \cite{AMdf}.
In spite of the sharp inequalities 
\begin{equation}
x_{n-1,l-1}^{(2)} < x_{nl}^{(1)} < x_{n+1,l+1},
\nonumber 
\end{equation}
which follow from the second of the rigorous sharp 
inequalities in Eq. (\ref{3p4p11}),
one soon finds that after a first few of initial zeros
[for instance for the Rabi model beginning with $l\gtrsim 2$ 
for $(\kappa,\Delta)=(0.2,0.4)$] \cite{AMtb}
\begin{equation}
 x_{n-1,l-1}^{(2)} \simeq x_{nl}^{(1)}  \simeq x_{n+1,l+1}.
\label{zrid}
\end{equation}
For $l\gtrsim 4$ and the Rabi model with 
$(\kappa,\Delta)=(0.2,0.4)$ the zeros 
coincide up to more than {\em five} decimal places 
(provided that $n$ is sufficiently large) - cf. data files \cite{AMdf}.
Because of the coagulation of zeros (\ref{zrid}),
the respective higher order poles and zeros of $\mathbb{F}(x)$
turn out soon to be closer to each other than machine precision. 
Thus any singularity and any zero of $\mathbb{F}(x)$, and most probably 
also that of $\mathbb{E}(x)$, become 
{\em numerically invisible} \cite{AMtb}. 
The latter implies that any practical 
implementation of the Schweber 
method that consists in locating zeros of $\mathbb{F}(x)$
{\em fails} for higher order eigenvalues \cite{AMtb}. 
Depending on model parameters, one can determine 
merely up to {\em 10}-{\em 20} eigenvalues, and that already 
in the exactly solvable limit of the displaced 
harmonic oscillator \cite{AMtb} - cf. F77 code made 
available online \cite{AMr}.

Our recipe for determining the first $N_0$ energy levels of 
$\hat{H}_b\in {\cal R}$ does not involve 
either searching for zeros of $\mathbb{F}(x)$ or for the poles of $\mathbb{E}(x)$
from the functional dependence of neither of the two functions.
Instead our analytic results enable one to get rid of 
both $\mathbb{F}(x)$ and $\mathbb{E}(x)$ and to focus exclusively
on the flows of polynomial zeros $x_{nl}$.
The recipe is as follows:

\begin{itemize}

\item Choose $N_c\ge N_0$ and determine the first $N_0$ zeros $x_{N_cl}$, $l\le N_0$, of
$P_{N_c}(x)$. Usually a good starting point is to take $N_c\approx N_0+20$. 
Because $P_{N_c}(x)$ has $N_c$ simple zeros, any omission of a zero
could be easily identified.

\item Gradually increase the cut-off value of $N_c$. The latter is what drives the 
incessant flows of polynomial zeros $x_{N_cl}$
[see the first sharp inequality in Eq. (\ref{1p5p4})], wherein each flow
is characterized by the parameter $l$.

\item Monitor convergence of the respective flows 
induced by the very first $n$ zeros of
$P_{N_c}(x)$. Each flow is a monotonically decreasing sequence  
having necessary a fixed limit point (\ref{1p5p6}).
Terminate your calculations when the $N_0$-th zero of 
$P_{N_c}(x)$ converged to $\xi_{N_0}$
within predetermined accuracy. Then as a rule all other 
flows $x_{N_cl}$ with $l<N_0$ have converged, too.

\end{itemize}
The examples of Ref. \cite{AMtb} show that the 
convergence of the zeros to the spectrum is very fast.
The numerical limits in calculating zeros 
were set by {\em over-} and {\em underflows}. 
Typically, with increasing $N_c$ the respective 
recurrences yielded first increasing and then decreasing 
$P_{N_c}(x)$. 
Here we sketch the basic features of an improved algorithm that allows 
to determine practically an {\em unlimited} number of energy levels 
within corresponding machine precision.
Our procedure to avoid the {\em over-} and {\em underflows} is rather
straightforward. Taking as an example the recurrences (\ref{rbmb2}) and 
common double precision, one monitors the magnitudes of 
the current three recurrence terms $\phi_{n+1}$, $\phi_{n}$, and $\phi_{n-1}$
as $n$ increases towards $N_c$. If the magnitude 
approaches $10^{308}$ ($10^{-308}$), 
the last three recurrence terms $\phi_{n+1}$, $\phi_{n}$, and $\phi_{n-1}$ are rescaled
by $10^{-308}$ ($10^{308}$). Because the recurrence 
coefficients are well behaving 
(they are fairly monotonic with exponents $\alpha=0$ and $\beta=-1$),
such a rescaling will move all three recurrence terms away from
{\em over-} or {\em underflows}.
The TTRR (\ref{rbmb2}) is then restarted anew
with the rescaled $\phi_{n+1}$, $\phi_{n}$, and $\phi_{n-1}$. 
Such a rescaling by a constant factor
obviously does not alter the position of zeros of the final $\phi_{N_c}$.
Also no loss of valid digits is involved, because
the change only involves exponent. Thus by the above rescaling one can
stitch the recurrence pieces together, thereby avoiding potential {\em over-} 
and {\em underflows}. The stitching can be continued up to the 
cut-off $N_c$ as large as the largest integer 
that can be stored within a given precision (e.g. $2^{53}$ in double precision
\cite{i3e}). Further numerical details are relegated to forthcoming 
publication \cite{mycpc}.

\section{Discussion}
\label{sec:disc}

\subsection{${\cal R}$ vs quasi-exactly-solvable models} 
\label{sec:dgn}
We have established the following properties of 
the models described by $\hat{H}_b\in {\cal R}$:
(i) the solution $\uppsi$ to the Schr\"{o}dinger 
equation (\ref{eme}) is the {\em generating function} for a set of polynomials 
$\{P_n(E)\}$ in the energy variable $E$, and (ii) 
the spectral points of $\hat{H}_b\in {\cal R}$ 
can be determined as fixed points of the flows 
generated by the polynomials zeros.
The properties resemble those of a subset of 
{\em quasi-exactly-solvable} (QES) problems of quantum 
mechanics \cite{TU,Trb,BD}. The QES models are distinguished
by the fact that a {\em finite} (and only a {\em finite}) part of 
their spectrum can be solved {\em analytically} and 
in {\em closed} form.
The corresponding energy eigenvalues are called 
the quasi-exact energy eigenvalues \cite{TU,Trb} and are commonly 
referred to as an {\em exceptional} spectrum \cite{Br}.
The solution to the Schr\"{o}dinger 
equation (\ref{eme}) is the {\em generating function} for a set of polynomials
in the energy variable $E$ \cite{BD} and
the quasi-exact energy eigenvalues can be determined
as the {\em zeros} of a critical Bender-Dunne polynomial $P_J(E)$ \cite{BD}.
The condition of quasi-exact solvability is 
reflected in the vanishing of the norm of 
all polynomials whose degree $n$ exceeds a 
critical value $J$ \cite{BD}. 
The corresponding moment functional ${\cal L}$ of such a polynomial
system is necessarily {\em degenerate} (cf. Appendix \ref{app:ttrr}).
Thus the Bender-Dunne polynomials do not form a conventional OPS.
Importantly, one speaks about the quasi-exact-solvability
already if the above properties apply for a discrete subset of 
model parameters.  
There are in general infinitely many parameters for which need not
exist any polynomial solution, yet a model is still called QES.

\subsection{Almost exactly solvable models}
\label{sec:aes}
The case when $\Uppsi\in\mathbb{C}^N$ 
can be {\em fully} diagonalized 
in the spin subspace (e.g. the Rabi model) 
deserves a special attention. A sufficient condition
for the full diagonalization is that the Hamiltonian $\hat{H}$ 
possesses an {\em Abelian} 
(e.g. cyclic) symmetry $G$ of the order $N$. 
Then $G$ has precisely $N$ {\em one-dimensional} 
irreducible representations (IR) $\Gamma_\gamma$, $\gamma= 1,2,\ldots, N$.
Let $g_j\in G$ are represented by $N\times N$ matrices 
$R_j$ in $\mathbb{C}^N$ (e.g. realized 
in terms of the {\em Sylvester} generator $S$ \cite{Al,Sa}).
The corresponding one-dimensional {\em orthogonal projectors} 
into particular IR $\Gamma_\gamma$ of $G$ are given by \cite{Mei,FiM}
\begin{equation}
P_\gamma  = (1/N)\, \sum_{j=1}^N  \chi_\gamma^*(g_j) R_j,
\nonumber 
\end{equation}
where $\chi_\gamma(g_j)$ are the characters of $g_j\in G$ 
in the given IR $\Gamma_\gamma$.
The total wave function $\Uppsi\in\mathbb{C}^N$
can be thus projected out into {\em one-dimensional} 
irreducible components, each 
satisfying its own eigenvalue equation (\ref{eme}). 
The FGT \cite{FG} employed for $N=2$ 
\cite{Br} can be considered as a special case of the more 
general {\em projection} method of 
explicitly determining irreducible representations of 
a {\em finite} group \cite{Mei,FiM}.
The property {\bf (c')} of Sec. \ref{sec:proof} of energy levels 
of the respective  $\hat{H}_b\in {\cal R}$ describing 
the one-dimensional irreducible components of $\Uppsi$ 
does not exclude {\em degeneracies} in the whole spectrum.
Any degeneracy corresponds to a nonzero 
overlap of the nondegenerate discrete spectra in the respective 
irreducible (e.g. parity invariant) subspaces governed 
by different $\hat{H}_b\in {\cal R}$ \cite{AMops}. 
In most cases the special points of the overlap correspond to the QES
part of the spectrum \cite{AMops,TAP}.

The {\em fully} diagonalizable models in the spin subspace  
could be thought of as 
{\em almost exactly solvable} (AES) models.
Indeed, had the flows of zeros terminated for some finite $N$, any such 
model would be considered as exactly, 
i.e. {\em algebraically}, solvable.
In contrast to the QES models, the {\em almost exact solvability} applies 
(i) to the entire spectrum and (ii) for all model parameters.
As exemplified by the QES Rabi model, the AES models
comprise some of the QES models.
Additionally, the example of a displaced harmonic oscillator shows that
the AES models may comprise {\em exactly solvable models}.

\subsection{Exactly solvable models}
\label{sec:es}
Let $\mathbb{D}_{x}$ be a suitable divided-difference operator
(discrete derivative) \cite{Ma88,Mg1,Wt} that maps $\Pi_{n}[x]$,
the linear space of polynomials in $x$ over $\mathbb{C}$ with
degree at most $n\in\mathbb{Z}_{\geq 0}$, into $\Pi_{n-1}[x]$
\cite{NSU,Lor,Ma88,Mg1,Wt,INS}.
Being a polynomial of degree $n-1$, one 
can represent $\mathbb{D}_{x} P_n(x)$ in general
only as (cf. Theorem I-2.2 of \cite{Chi,Ma88,Wt})
\begin{equation}
\mathbb{D}_{x} P_n(x) = \sum_{r=0}^{n-1} c_{n,r}P_r(x),
\nonumber 
\end{equation}
with some constant coefficients $c_{n,r}$. 
The hallmark of exactly solvable models is existence
of a {\em structure relation} satisfied by the 
corresponding OPS $\{P_n(x)\}_{n=0}^\infty$,
\begin{equation}
\mathbb{D}_{x} P_n(x) = -B_n(x) P_n(x) + A_n(x) P_{n-1}(x),
\label{pnlm}
\end{equation}
where the coefficients 
$A_n(x)$ and $B_n(x)$ are in general nonpolynomial 
functions \cite{NSU,Lor,Ma88,Mg1,Wt,INS}. 
Obviously, if there is one structure relation 
(\ref{pnlm}), there is another one. 
The other one results by expressing $P_{n-1}$ from 
the fundamental TTRR (\ref{chi3tr}) and substituting it back 
into the original structure relation (\ref{pnlm}).
The resulting pair of structure relations 
(i) leads directly to a pair of mutually adjoint raising and lowering 
{\em ladder operators} \cite{INS}, 
(ii) implies that orthogonal polynomials satisfy in general 
a second-order difference equation (cf. Sec. 4 of Ref. \cite{INS}), and 
(iii) allows one to introduce a {\em discrete} analogue 
of the {\em Bethe Ansatz} equations
(cf. Sec. 5 of Ref. \cite{INS}).
The structure relation (\ref{pnlm})
can be established for any {\em classical} OPS 
(p. 783 of Ref. \cite{AS}; Section 6 of Ref. \cite{AlS};
Proposition 2.6 of Ref. \cite{GMS}),
{\em semi-classical} OPS (Theorem 1 of Ref. \cite{Mg1}; 
Proposition 4.4 of Ref. \cite{Wt}), and any 
OPS orthogonal with respect to a 
discrete measure supported on {\em equidistant} points 
(Theorem 1.1 of Ref. \cite{INS}).
[In the {\em semi-classical} case, the function
$\mathbb{E}(x)$ itself satisfies a first order difference equation
with polynomial coefficients (Theorem 1 of Ref. \cite{Mg1}; 
Proposition 4.1 of Ref. \cite{Wt}).]
In brief one finds a structure relation only for the OPS which belong 
to the Askey scheme (p. 183 of Ref. \cite{KLS})
or to the $q$-analogue of the Askey scheme (p. 413 of Ref. \cite{KLS}).
In each of the above cases, 
$\Lambda\equiv \Xi$ representing the infinite discrete 
support of $d\nu(x)$ is necessarily one 
of four primary classes of special non-uniform lattices \cite{Ma88,Mg1,Wt}:
the linear lattice, the linear $q$-lattice, the quadratic lattice, and
the $q$-quadratic lattice (for their properties see Table 2 of Ref. \cite{Wt}). 
The $q$-quadratic lattice, in its general non-symmetrical form, 
is the most general case and the other lattices can be found 
from this by limiting processes \cite{Wt}.
More specifically, either
\begin{equation}
\Lambda=\{x \,| \,x=u_2n^2+u_1n+u_0,\, n\in\mathbb{N} \},
\label{tpq}
\end{equation}
or 
\begin{equation}
\Lambda=\{x \,| \,x=u_2q^{-n}+u_1q^n+u_0,\, n\in\mathbb{N} \},
\label{tpqq}
\end{equation}
where $u_j$ are real constants and $0<q<1$.
Thus the spectrum $\Sigma$ of an exactly solvable $H_b\in {\cal R}$
can only assume one of the above forms of $\Lambda$.
Thereby we have independently arrived at essentially the same type
of exactly solvable spectra as did Odake and Sasaki \cite{OS8} 
within the framework of their exactly solvable discrete quantum mechanics
with real shifts (cf. Eqs. (4.7-11) of Ref. \cite{OS8}).
Conversely, it appears that if $\Sigma\ne \Lambda$, where $\Lambda$ is 
one of the above types (\ref{tpq}) and (\ref{tpqq}), 
then the model {\em cannot} be exactly solvable.

An example is provided by exactly solvable displaced harmonic oscillator
having equidistant spectrum $\Upsigma$.
The corresponding OPS is that of the Charlier polynomials \cite{Chi,Chr} 
and $\Lambda\equiv\Xi$ is an equidistant lattice
which coincides with $\Upsigma$ \cite{AMtb}. 
$\Delta\ne 0$ in the Rabi Hamiltonian (\ref{rabih}) induces a deformation of 
the Charlier polynomials to {\em non-classical} discrete
orthogonal polynomials and, at the same time, a 
deformation of the underlying equidistant lattice $\Lambda$. 
The deformed lattice does not correspond to any 
of the primary lattice classes implying that the Rabi model 
is not exactly solvable. Although neither the weight function
nor the deformed lattice are analytically known, 
the above deformation is a {\em norm preserving deformation} of the
underlying OPS \cite{AMtb}.

{\em Algorithmic complexity} theory \cite{PML} has been used, 
although without much success, to discuss the degree of 
randomness of the sequence of energy 
eigenvalues of conservative quantum systems \cite{GGV}.
In the present case, an alignment of the physical spectrum $\Upsigma$
with one of the primary classes (\ref{tpq}) and (\ref{tpqq}) 
of lattices $\Lambda\equiv \Xi$
can identify a model as exactly solvable. 
The degree of randomness of the sequence of energy eigenvalues could 
be then defined as a minimal distance from the four primary classes of 
special non-uniform lattices. Further details will be discussed
elsewhere \cite{mycpc}.

\subsection{Comparison with other numerical methods}
\label{sec:dnm}
Our method of determining energy levels 
differs from any of the known methods that involve 
(i) a brute force numerical diagonalization, 
(ii) computation of a correlation function 
$\langle\uppsi({\bf r},0)| \uppsi({\bf r},t)\rangle$ 
from a numerical solution $\uppsi({\bf r},t)$
as in a spectral method by Feit et al \cite{FFS},
(iii) searching for zeros of analytic functions having infinite number 
of poles and zeros on the real axis (e.g. determined by infinite continued
fractions as in the Schweber method 
(cf. Eq. (A.16) of Ref. \cite{Schw}) or as in Braak's approach \cite{Br}), and
(iv) numerical diagonalization using Hill's determinant approach \cite{WL}.

A brute force numerical diagonalization allows one 
to determine around {\em 2000} energy levels 
in double precision (ca 16 digits) for the Rabi model. 
This is much less than is possible by our approach.
Also any deeper analytic insight is missing.  
Searching for zeros of analytic functions yields 
only ca. {\em 20} levels. Employing further tricks one can hardly 
overcome the range of $\sim 100$ levels.
Using Hill's determinant approach one can determine $\sim 500$ levels,
which is still merely half of what was possible to obtain by 
the simple stepping algorithm employed in Ref. \cite{AMtb}.

\subsection{Relation with earlier work} 
\label{sec:drl}
A proof of our main result in the special case of a displaced harmonic 
oscillator has been provided in our earlier work \cite{AMtb}. 
Yet the proof was not general.
It was made possible thanks to a largely fortuitous coincidence
that the orthogonal polynomials
of discrete variable relevant for the displaced harmonic 
oscillator are the well-known
(monic) {\em Charlier polynomials} \cite{Chr} 
(cf. Eqs. VI-1.4-5 of Ref. \cite{Chi}) \cite{AMtb}.
For the case of the Rabi model \cite{Schw,Rb}, the relevant 
orthogonal polynomials of discrete variable are not classical one
and have not been studied in detail so far \cite{AMtb}. As a consequence,  
only a partial proof of the above statement could have 
been provided that was limited to the case when 
a dimensionless interaction constant $\kappa< 1$ \cite{AMtb}.
On using the identity (\ref{ttdf}), the latter has now been 
proven rigorously, thereby confirming earlier numerical
evidence that the statement remains to be valid 
also for $\kappa\ge 1$ \cite{AMtb}.

On adopting the notation $\psi(n)=p_n(x)$,
our TTRR (\ref{chi3tr}) can be interpreted as 
a {\em finite difference} Schr\"odinger equation \cite{OS8,SVZl}
\begin{equation}
\psi(n+1) + c_{n}\psi(n) + \lambda_{n} \psi(n-1) = x\psi(n),
\label{chi3trdm}
\end{equation}
with $x$ playing the role of an {\em eigenvalue}.
Thus our {\em recurrence class of 
purely bosonic models with polynomial coefficients}, ${\cal R}$,
provides a realization of $\widetilde{\mathcal H}$ of Odake and Sasaki
in the so-called discrete quantum mechanics (dQM) 
with {\em real shifts} (rdQM) (cf. Sec. III of Ref. \cite{OS8}). 
Eq. (\ref{chi3trdm}) has been earlier studied by Spiridonov et al \cite{SVZl}.

\section{Conclusions}
\label{sec:conc}
A class ${\cal R}$ of purely bosonic models has been characterized 
having the following properties in the Bargmann Hilbert space 
of analytic functions: 
(i) wave function 
$\uppsi(\upepsilon,z)=\sum_{n=0}^\infty \phi_n(\upepsilon) z^n$ 
is the {\em generating function} 
for orthogonal polynomials $\phi_n(\upepsilon)$ of 
a {\em discrete} energy variable $\upepsilon$, 
(ii) 
any Hamiltonian $\hat{H}_b\in {\cal R}$ 
has nondegenerate purely point spectrum that corresponds 
to infinite discrete support of measure
$d\nu(x)$ in the orthogonality relation of the polynomials 
$\phi_n$,
(iii) the support is determined exclusively by
the points of discontinuity of $\nu(x)$,
(iv) the spectrum of $\hat{H}_b\in {\cal R}$ can be numerically
determined as fixed points of monotonic flows of the zeros of 
orthogonal polynomials $\phi_n(\upepsilon)$,
(v) one can compute practically an unlimited number of 
energy levels (e.g. $2^{53}$ in double precision).
If a model of ${\cal R}$ is exactly solvable, its spectrum 
can only assume one of four qualitatively different types.
Our results were shown to apply to a class of spin-boson quantum models 
that are, at least partially, diagonalizable in a spin subspace.
The class is broad enough to encompass the Rabi model 
and its various generalizations.

\section{Acknowledgment}
I thank Prof.'s W. Gautschi, M. E. H. Ismail, N. S. Witte,
Y.-Z. Zhang, and A. Zhedanov for discussion.
Continuous support of MAKM is largely acknowledged.

\appendix

\section{Mathematical remarks}
\label{app:tchr}

\subsection{Moment functional ${\cal L}$}
\label{app:ttrr}
Satisfying the TTRR such as (\ref{chi3tr}) is the necessary 
and sufficient condition that
there exists a unique positive definite moment functional
${\cal L}$, such that
for the family of polynomials $\{p_n\}$ holds
\begin{equation}
{\cal L}[1]=\lambda_0,~~~~~
  {\cal L}[p_m(x)p_n(x)]= \lambda_0\lambda_1\ldots \lambda_{n}\delta_{mn},
\end{equation}
where $m,n=0,1,2,\ldots$ and $\delta_{mn}$ is the Kronecker symbol. 
Thereby the polynomials $\{p_n\}$ form an OPS. 
Because $\lambda_n > 0$, the norm 
of the polynomials $p_n$ is positive definite, 
${\cal L}[p_n^2(x)]>0$, and ${\cal L}$ 
is a {\em positive definite nondegenerate} moment functional 
(p. 16 of Ref. \cite{Chi}). It is reminded here that the 
QES models have {\em degenerate} ${\cal L}$, i. e., 
$\lambda_{J+1}=0$ for some $J>0$.

According to the {\em representation theorem} (Theorem II-3.1 of Ref. \cite{Chi}), 
the {\em distribution} function $\nu$ of the positive moment functional 
${\cal L}$,
\begin{equation}
{\cal L}[x^n]=\int_{-\infty}^\infty x^n\, d\nu(x)
                       =\mu_n ~~~~~~~~(n=0,1,\ldots),
\label{mf}
\end{equation}
is the limit of a sequence of bounded, right continuous, 
nondecreasing step functions $\nu_n(x)$'s, 
\begin{eqnarray}
\nu_n(x) &=& 0~~~~~~~~~ (-\infty\le x < x_{n1}), 
\nonumber\\
\nu_n(x)&=& M_{n1}+\ldots+ M_{np} ~~(x_{np} \le x < x_{n,p+1}), 
\nonumber\\
\nu_n(x)&=& \mu_0~~~~~~~ (x \ge x_{nn}),
\label{psin}
\end{eqnarray}
where $x_{nl}$, $l=1,2,\ldots,n$, are the zeros of $p_n(x)$, 
Consequently
\begin{itemize}

\item $\nu_n(x)$ has exactly $n$ points 
of increase, $x_{nk}$,

\item the discontinuity of $\nu_n(x)$ at each $x_{nk}$  
equals $M_{nk}$ ($k = 1, 2, \ldots, n$),

\item at least the first $(2n-1)$ moments of $\nu_n(x)$ are identical 
with those of $\nu(x)$, i.e.,
\begin{equation}
\int_{-\infty}^\infty x^l\,d\nu_n(x)=\mu_l 
~~~~~~~~~~(l = 0,1, 2, \ldots, 2n-1).
\end{equation}

\end{itemize}

\subsection{The ratio in Eq. (\ref{Exe}) for a finite $n$}
\label{app:gen}
An indication of that the poles of $\mathbb{E}(x)$ could correspond 
to the set $\Xi$ is provided by considering the 
ratio in (\ref{Exe}) for a finite $n$.
Then the ratio in (\ref{Exe}) enables the partial 
fraction decomposition  
(Theorem III-4.3 of Ref. \cite{Chi}),
\begin{equation}
\frac{P_{n-1}^{(1)}(z)}{P_n(z)} 
   =\int_{-\infty}^\infty \frac{d\nu_n(x)}{z-x}
     =\sum_{k=1}^n \frac{M_{nk}}{z-x_{nk}},
\label{pfdrl}
\end{equation}
where the numbers $M_{nl}$ are all {\em positive}
(cf. Appendix \ref{app:ttrr})
and satisfy the condition $\sum_{l=1}^n M_{nl}=1$ \cite{AMops,Chi}.

\subsection{Further consequences of the Perron-Kreuser theorem}
\label{app:pkth}
The Perron-Kreuser theorem (Theorem 2.3 in Ref. \cite{Gt}) implies that
the dominant solutions of the TTRR (\ref{3trg}) do not 
generate an element of $\mathfrak{b}$ if
\begin{itemize}

\item {\bf (a'')} $2\alpha>\beta$ and either 
         {\bf (i)} $\alpha> -1/2$ 
                  or {\bf (ii)} $\alpha=-1/2$ and $|a|\ge 1$,

\item {\bf (b'')} $2\alpha=\beta$ and either 
         {\bf (i)} $\alpha> -1/2$ 
            or {\bf (ii)} $\alpha=-1/2$ and the larger root 
               $|t_2|\le|t_1|$ of $t^2+at+b=0$ satisfies $|t_1|\ge 1$.

\end{itemize}

\subsection{Stieltjes transform}
\label{app:strf}
One can, in principle, find $\nu$ and determine 
its infinite discrete support $\Lambda\equiv \Xi$ by inverting the 
Stieltjes transform (\ref{erpt}). 
Indeed if the representation (\ref{erpt}) holds for $z\not\in \mathbb{R}$, 
then \cite{ST,Ak}
\begin{eqnarray}
\lefteqn{
\frac{1}{2}\, [\nu(x_2)+\nu(x_2-0)] - \frac{1}{2}\, [\nu(x_1)+\nu(x_1-0)]
}
\nonumber
\\
&& = - \frac{1}{2\pi i} \lim_{\epsilon\rightarrow 0_+} 
\int_{x_1}^{x_2} 
[\mathbb{E}(t+i\epsilon) - \mathbb{E}(t-i\epsilon)]\, dt.
\end{eqnarray}
Recovering $\nu$ from $\mathbb{E}$ constitutes the famous 
{\em problem of moments} \cite{ST,Ak}.
Every isolated pole $z=u$ of $\mathbb{E}(z)$ contributes 
a discrete mass of $\nu$ at $x = u$ and the mass equals 
the residue of $\mathbb{E}(z)$ at $z=u$, which 
is given by Eq. (\ref{rsfrme}). 
At all other points the limit in our case vanishes.
The orthogonality measures of several important systems of orthogonal
polynomials were found by (i) computing the large $n$ asymptotic of $P_n(x)$ and
$P_n^{(1)}(x)$ in the representation (\ref{erpt}) followed by
(ii) the inversion of the Stieltjes transform \cite{Isb}.




\begin{thebibliography}{99}
\bibitem{ntt}
The reason for a different nomenclature regarding TTRR's 
(\ref{3trg}) and (\ref{chi3tr}) is that the
TTRR  of the form (\ref{3trg}) is common in general theory 
of recurrences \cite{Gt}, whereas the TTRR of the form (\ref{chi3tr}) 
are employed in the theory of orthogonal polynomials \cite{Chi}.
We maintained the respective notations in order to facilitate comparison
with relevant results of Refs. \cite{Gt,Chi}.
\end{thebibliography}
\end{document}